\documentclass[onecolumn,aps,prd,superscriptaddress,preprintnumbers,nofootinbib,10pt]{revtex4-2}
\usepackage{amsfonts}
\usepackage[tbtags]{amsmath}
\usepackage{amssymb}
\usepackage{color}
\usepackage{courier}
\usepackage{dsfont}
\usepackage{euscript}
\usepackage{epsfig}
\usepackage{epstopdf}
\usepackage{graphics}
\usepackage{graphicx}
\usepackage[utf8]{inputenc}
\usepackage{latexsym}
\usepackage{MnSymbol}
\usepackage{pifont}
\usepackage{physics}
\usepackage{slashed}
\usepackage{xcolor}
\usepackage[normalem]{ulem}
\usepackage{url}
\usepackage{verbatim}
\usepackage{widetable}
\usepackage{setspace}
\usepackage[plainpages=false,pdfpagelabels]{hyperref}

\allowdisplaybreaks

\hypersetup{
	colorlinks=true,
	citecolor=blue,
	citebordercolor=red,
	linktoc=all,
	linkcolor=blue,
	urlcolor=blue
}

\begin{document}

		\title{\texorpdfstring{\bf Bosonization, vertex operators and maximal violation of \\ the Bell-CHSH inequality in wedge regions}{Bosonization, vertex operators and maximal violation of the Bell-CHSH inequality in wedge regions}}

	\author{J. G.A. Caribé}\email{joaogcaribe@ufrrj.br} \affiliation{ UFRRJ, Universidade Federal Rural do Rio de Janeiro, Departamento de Física, Zona Rural, BR-465, Km 07, 23890-000, Seropédica, Rio de Janeiro, Brazil}\affiliation{UERJ -- Universidade do Estado do Rio de Janeiro,	Instituto de Física -- Departamento de Física Teórica -- Rua São Francisco Xavier 524, 20550-013, Maracanã, Rio de Janeiro, Brazil}

	\author{M. S.  Guimaraes}\email{msguimaraes@uerj.br} \affiliation{UERJ -- Universidade do Estado do Rio de Janeiro,	Instituto de Física -- Departamento de Física Teórica -- Rua São Francisco Xavier 524, 20550-013, Maracanã, Rio de Janeiro, Brazil}
	
	\author{I. Roditi} \email{roditi@cbpf.br} \affiliation{CBPF $-$ Centro Brasileiro de Pesquisas Físicas, Rua Dr. Xavier Sigaud 150, 22290-180, Rio de Janeiro, Brazil}\affiliation{Institute for Theoretical Physics, ETH Z\"urich, 8093 Z\"urich, Switzerland} 
	
	\author{S. P. Sorella} \email{silvio.sorella@fis.uerj.br} \affiliation{UERJ -- Universidade do Estado do Rio de Janeiro,	Instituto de Física -- Departamento de Física Teórica -- Rua São Francisco Xavier 524, 20550-013, Maracanã, Rio de Janeiro, Brazil}

	\begin{abstract}

	It is pointed out that the vertex operators of a chiral boson in $1+1$ dimensions provide an explicit realization of dichotomic, bounded, Hermitian operators that saturate the Tsirelson bound of the Bell-CHSH inequality in the vacuum state.

\end{abstract}
		
\maketitle

\section{Introduction}\label{sect}

Since the seminal work of Summers and Werner \cite{Summers:1987fn,Summers:1987squ,Summers:1987ze}, the study of the Bell-CHSH inequality \cite{Bell:1964kc,Clauser:1969ny} in relativistic Quantum Field Theory has attracted increasing interest. For recent overviews, see \cite{Guimaraes:2024byw,Guimaraes:2024mmp}. The modern program combines the Haag-Kastler framework, the Bisognano-Wichmann theorem, and Tomita-Takesaki modular theory \cite{Haag:1992hx,Bisognano:1975ih,Takesaki:1970aki,Bratteli:1979tw,Summers:2003tf,Guido:2008jk}. In this setting, explicit Bell-CHSH constructions are already known for free fields, including scalar realizations in terms of bounded unitary or Hermitian operators \cite{Guimaraes:2024alk,Guimaraes:2025vfu} and maximal-violation constructions for massless spinors \cite{Dudal:2023mij}.

The Summers-Werner theorem states that the Bell-CHSH correlator in wedge regions can be made arbitrarily close to the Tsirelson bound in the vacuum of a free Quantum Field Theory \cite{Summers:1987fn,Summers:1987squ,Summers:1987ze}. Thus, for suitable wedge-supported test functions, one can construct a set $\{A(f),A(f'),A(g),A(g')\}$ of dichotomic\footnote{A smeared operator $A(f)$ is said to be dichotomic when it satisfies $A^2(f)=1$.}, bounded, Hermitian operators such that the vacuum correlator\footnote{The quantities $(f,f')$ and $(g,g')$ denote pairs of smooth space-like test functions needed to properly define $\{A\}$ as operators acting on the 1-particle Hilbert space of the theory.}
\begin{equation} 
\langle {\cal C} \rangle = \langle 0|\; (A(f) + A(f'))A(g) + (A(f)-A(f'))A(g') \;|0\rangle \;, \label{corr}
\end{equation}
can be made arbitrarily close to Tsirelson's bound \cite{Cirelson:1980ry}, namely 
\begin{equation} 
\langle {\cal C} \rangle \approx 2\sqrt{2} \;. \label{Ts}
\end{equation}

Despite the generality of the theorem, the Fermi and Bose fields behave differently when one seeks explicit Bell operators. On the fermionic side, the canonical anti-commutation relations allow one to write down the relevant dichotomic, bounded, Hermitian operators directly. For example, in the case of a massless Majorana spinor \cite{Dudal:2023mij}
\begin{equation}
\psi(f) = \int_{-\infty}^\infty dx_{+}\; f_R(x_+) \psi_R(x_+) + \int_{-\infty}^{\infty} dx_-\; f_L(x_-) \psi_L(x_-) \;, \label{mj}
\end{equation}
where $\psi_R(x_+)=\psi_R(t+x)$ and $\psi_L(x_-)=\psi_L(t-x)$ stand, respectively, for the real-valued right- and left-moving components, the operator 
\begin{equation} 
A_M(f) = \frac{\psi(f)}{||f||} \label{dm}
\end{equation} 
is  bounded, Hermitian and dichotomic for any two-component real-valued smooth test function, $f=(f_R(x_+), f_L(x_-))$. By suitably choosing a set  $\{f,f',g,g'\}$ of four test functions, which play a role akin to the four Bell's angles of Quantum Mechanics, one can define a Bell-CHSH correlator that saturates Tsirelson's bound \cite{Summers:1987fn,Summers:1987squ,Summers:1987ze}. In the present example, the inner product between two test-functions is
\begin{equation} 
\langle f | g\rangle = \int_0^\infty dk f_R(k)^* g_R(k) + \int_{-\infty}^0 dk f_L(k)^* g_L(k)  \;, \label{inn}
\end{equation}
where
 $f_R(k)$ and $f_L(k)$ are, respectively, the Fourier transformations of $f_R(x_+)$ and $f_L(x_-)$
\begin{equation} 
f_R(k) = \int dx_+ e^{i kx_+} f_R(x_+) \;, \qquad f_L(k) = \int dx_- e^{i kx_-} f_L(x_-) \;. \label{ft}
\end{equation}
It turns out that, for the operator \eqref{dm}, 
\begin{equation} 
\langle 0| A_M(f) A_M(g) |0\rangle = - i \frac{\langle f|g\rangle}{||f||.||g||} \;, \label{Mcorr}
\end{equation}
so that, for the Fermi field case, the Bell-CHSH correlator reads 
\begin{equation} 
\langle C \rangle = - i \left( \frac{\langle f|g\rangle}{||f||.||g||}+  \frac{\langle f'|g\rangle}{||f'||.||g||}+\frac{\langle f|g'\rangle}{||f||.||g'||} -\frac{\langle f'|g'\rangle}{||f'||.||g'||}  \right) \;. \label{MB}
\end{equation} 
At this stage, one can follow two equivalent routes to suitably choose the test functions $\{f,f',g,g'\}$. In the first, one employs the original Summers-Werner construction based on the Bisognano-Wichmann theorem and Tomita-Takesaki modular theory \cite{Summers:1987fn,Summers:1987squ,Summers:1987ze}. In the second, developed recently in \cite{Dudal:2026eil}, one relates the inner products in \eqref{MB} to the spectral properties of Carleman and Hankel operators. In both cases, maximal violation is achieved. \\\\For Bose fields, explicit examples of Bell operators are already known by making use of the Weyl unitaries \cite{Guimaraes:2024alk,Guimaraes:2025vfu}, although not leading to maximal violation.  The narrower question addressed here is whether one can realize the same Bell maximal violating  mechanism  exhibited by Fermi fields  directly within the  framework of the bosonization  of chiral fields. The aim of the present letter is to show that the answer is affirmative: by relying on the bosonization of $(1+1)$-dimensional models \cite{Coleman:1985rnk,Senechal:1999us}, we obtain a vertex-operator realization of the fermionic Bell construction. More precisely, for the bosonization value $\alpha^2=4\pi$, Hermitian smeared combinations of the chiral vertex operators reproduce the two-point function and normalization entering the Majorana Bell operators. There is no need to rederive the modular choice of wedge-supported test functions; once the bosonized operator family is identified, the near-Tsirelson result follows from the same Summers-Werner input as in the fermionic case  \cite{Summers:1987fn,Summers:1987squ,Summers:1987ze}.

\section{Vertex operators and the saturation of the Tsirelson bound }\label{Bose}
	
One starts by considering a free massless scalar field in $(1+1)$-dimensional Minkowski spacetime. As one learns from the bosonization setup  \cite{Coleman:1985rnk,Senechal:1999us}, the scalar field splits into right- and left-movers $\varphi_R(x_+)$ and $\varphi_L(x_-)$, obeying
\begin{equation} 
\partial_- \varphi_R(x_+) = 0 \;, \qquad \partial_+ \varphi_L(x_-) = 0  \;. \label{movers}
\end{equation}
Our strategy is simple. First, at the bosonization value $\alpha^2=4\pi$, introduce conjugate vertex operators satisfying a fermionic exchange relation. Second, construct Hermitian smeared combinations of these operators reproducing the two-point function entering the Majorana Bell construction. Finally, one uses the same wedge-supported test functions as in the fermionic case, so that the localization in causal complementary wedge algebras and the modular optimization of the CHSH correlator are inherited directly from the Summers-Werner framework \cite{Summers:1987fn,Summers:1987squ,Summers:1987ze}.

The relevant operators of the theory as, for instance, the chiral currents $(J_+,J_-)$, are obtained through the right- and left-movers. A particular class of operators, called vertex operators, is obtained by exponentiating the left- and right-movers and taking the normal ordering, namely 
\begin{equation} 
V_R^{\alpha}(x_+) = \;: e^{i \alpha \varphi_R(x_+) } : \;, \label{vop}
\end{equation}
where $\alpha$ is a free parameter. An analogous expression holds for the left vertex operator $V_L^{\alpha}(x_-)$. A useful feature of the vertex operators is the exchange relation
\begin{equation} 
V_R^{\alpha}(x_+) V_R^{\beta}(x_+') = e^{\frac{i \alpha \beta}{4} sign(x_+ -x_+')}\;V_R^{\beta}(x_+') V_R^{\alpha}(x_+) \;, \label{ccv}
\end{equation}
For conjugate charges $\beta=-\alpha$ and $\alpha^2=4\pi$, the phase factor equals $-1$ for $x_+\neq x_+'$, so the pair $V_R^{\alpha}$ and $V_R^{-\alpha}$ behaves as a pair of fermions. Needless to say, the vertex operators \eqref{vop} are a fundamental tool in order to map Fermi theories into Bose ones and vice-versa. Furthermore,  
\begin{equation} 
\langle 0| V_R^{\alpha}(x_+) V_R^{-\alpha}(x_+') |0\rangle = \frac{1}{x_+-x_+'- i \varepsilon} \;, \qquad \alpha^2 = 4 \pi \;, \label{corrv}
\end{equation}
which coincides with the two-point function of a chiral massless Fermi field. To approach the Bell-CHSH inequality, we introduce the Hermitian, smeared, vertex operators 
\begin{equation} 
Q_R^{\alpha}(f) = \frac{1}{\sqrt{2}} \int_{-\infty}^\infty dx_+ f(x_+) \left( V_R^{\alpha}(x_+) + V_R^{\alpha}(x_+)^\dagger \right)  \;, \label{qv}
\end{equation}
where $f$ is a real-valued smooth test function. It follows that 
\begin{equation} 
\langle 0| Q_R^{\alpha}(f) Q_R^{\alpha}(g) |0\rangle = \frac{1}{2} \int dx_+ dx_+' f(x_+) g(x_+') \frac{1}{x_+-x_+'- i \varepsilon} = -i \int_0^\infty dk f(k)^* g(k)  \;, \qquad \alpha^2=4 \pi \;. \label{scv}
\end{equation}
Thus, at the level relevant for the Bell construction, $Q_R^\alpha(f)$ reproduces the chiral Majorana two-point function. Adding now the left component, one obtains a full Hermitian operator whose vacuum two-point function matches the massless Hermitian Majorana operator \eqref{mj}, namely 
\begin{equation} 
{\hat W^{\alpha}}(f) =  \frac{1}{\sqrt{2}} \int_{-\infty}^\infty dx_+ f_R(x_+) \left( V_R^{\alpha}(x_+) + V_R^{\alpha}(x_+)^\dagger \right)  +  \frac{1}{\sqrt{2}} \int_{-\infty}^\infty dx_- f_L(x_-) \left( V_L^{\alpha}(x_-) + V_L^{\alpha}(x_-)^\dagger \right)  \;, \label{wv}
\end{equation}
with 
\begin{equation} 
\langle0| {\hat W^{\alpha}}(f) {\hat W^{\alpha}}(g)|0  \rangle = -i \langle f|g\rangle = -i \left( \int_0^\infty dk f_R(k)^* g_R(k) + \int_{-\infty}^0 dk f_L(k)^* g_L(k) \right) \;, \label{binn}
\end{equation}
which is exactly the inner product of expression \eqref{inn}. In this sense, the bosonization dictionary identifies ${\hat W^\alpha}(f)$ with the bosonic realization of the Majorana Bell operator. It follows that the Bell-CHSH correlator built from vertex operators, namely 
\begin{equation} 
W^{\alpha}(f) = \frac{{\hat W^{\alpha}}(f) }{||f||} \;, \label{wfv}
\end{equation}
\begin{eqnarray} 
\langle C\rangle_{vertex} & = & \langle 0| (W^{\alpha}(f)+W^{\alpha}(f'))W^{\alpha}(g)+(W^{\alpha}(f)-W^{\alpha}(f'))W^{\alpha}(g') |0\rangle  \nonumber \\
&=& - i \left( \frac{\langle f|g\rangle}{||f||.||g||}+  \frac{\langle f'|g\rangle}{||f'||.||g||}+\frac{\langle f|g'\rangle}{||f||.||g'||} -\frac{\langle f'|g'\rangle}{||f'||.||g'||}  \right) \;, \label{VB}
\end{eqnarray} 
can therefore be made arbitrarily close to Tsirelson's bound:
\begin{equation} 
\langle C\rangle_{vertex} \approx 2 \sqrt{2} \;. \label{satv}
\end{equation}
Indeed, from \cite{Summers:1987fn,Summers:1987squ,Summers:1987ze}, it follows that, by using the Bisognano-Wichmann results in wedge regions and the Tomita-Takesaki modular theory, the inner products can be chosen in such a way that 
\begin{equation} 
\frac{\langle f|g\rangle}{||f||.||g||}=  \frac{\langle f'|g\rangle}{||f'||.||g||}=\frac{\langle f|g'\rangle}{||f||.||g'||} =-\frac{\langle f'|g'\rangle}{||f'||.||g'||} = i \sqrt{2} \frac{\lambda}{1+\lambda^2} \;, \label{lv}
\end{equation}
where $\lambda \in (0,1)$ is the spectral parameter of the modular operator $\delta$ \cite{Summers:1987fn,Summers:1987squ,Summers:1987ze} 
\begin{equation} 
\delta = e^{- 2\pi K} \;, \label{tt} 
\end{equation}
with $K$ being the boost generator \cite{Summers:1987fn,Summers:1987squ,Summers:1987ze}. Thus 
\begin{equation} 
\langle C\rangle_{vertex} = 2\sqrt{2} \frac{2 \lambda}{1+\lambda^2}  \underset{\lambda \rightarrow 1} \rightarrow 2 \sqrt{2} \;. \label{fvv}
\end{equation}

\section*{Acknowledgments}
The authors would like to thank the Brazilian agencies CNPq, CAPES and FAPERJ for financial support.  S. P.~Sorella, I.~Roditi, and M. S.~Guimaraes are CNPq researchers under contracts 302991/2024-7, 311876/2021-8, and 309793/2023-8, respectively. Prof. Ricardo Correa da Silva is gratefully acknowledged for fruitful discussion. 


\end{document}